\documentclass[journal]{IEEEtran}
\IEEEoverridecommandlockouts
\usepackage{cite}

\usepackage{amsmath,amssymb,amsfonts,mathtools,bm}
\usepackage{amsthm}
\usepackage{algorithm}
\usepackage{algorithmic}
\usepackage{graphicx}
\usepackage{textcomp}
\usepackage{xcolor,float}
\usepackage{tabularx}
\usepackage{textcomp}
\usepackage{diagbox}
\usepackage{subfig}
\usepackage{graphicx} 
\usepackage{subcaption}
\usepackage{svg}
\usepackage{booktabs}
\usepackage{subcaption} 
\usepackage{pifont}
\usepackage{hyperref}
\usepackage{multirow}

\usepackage{adjustbox}

\allowdisplaybreaks
\renewcommand{\baselinestretch}{1}

\begin{document}

\title{RadioDiff-Loc: Diffusion Model Enhanced Scattering Congnition for NLoS Localization with Sparse Radio Map Estimation}
\author{
Xiucheng Wang,
Qiming Zhang,
Nan Cheng

\thanks{ }
\thanks{\par Xiucheng Wang and Nan Cheng are with the School of Telecommunication Engineering in Xidian University.
\par Qiming Zhang is with the School of Artificial Intelligence in Xidian University.
}

}
    
    \maketitle

\IEEEdisplaynontitleabstractindextext

\IEEEpeerreviewmaketitle

\begin{abstract}
Accurate localization of non-cooperative signal sources in non-line-of-sight (NLoS) environments is a fundamental yet unsolved challenge, with critical implications for autonomous navigation, emergency response, and smart infrastructure. Traditional localization techniques—relying on line-of-sight (LoS) paths, structured pilots, or known transmit power—fail in environments characterized by severe multipath propagation, restricted sensing regions, and passive emitters. To address this, we propose RadioDiff-Loc, a novel generative localization framework that integrates conditional denoising diffusion models with physics-guided sampling strategies. The model is first trained to learn the statistical prior of the radio map (RM) conditioned on environmental geometry, and then reconstructs the full signal field from a sparse set of received signal strength (RSS) observations. Localization is performed by identifying the intensity peak of the generated RM. To overcome the lack of transmit power knowledge, we introduce a power-invariant normalization scheme, enabling robust inference from uncalibrated RSS. Furthermore, inspired by knife-edge diffraction theory, we design two geometry-aware sampling strategies—surface-based and vertex-based—that place sensors at locations with maximal information gain. These methods drastically reduce measurement costs by aligning sampling complexity with environmental geometry, not emitter position. Extensive experiments demonstrate that RadioDiff-Loc achieves high localization accuracy using over 10× fewer samples than baseline approaches, offering a scalable, interpretable, and physically grounded solution for non-cooperative localization under NLoS conditions.
\end{abstract}


\section{Introduction}
The rapid growth of autonomous systems, Internet of Things (IoT) deployments, and smart infrastructure in urban and industrial environments has sharply increased the demand for reliable localization under complex and obstructed conditions \cite{6g,lin2016enhanced,li2020toward}. Applications such as indoor navigation in large buildings \cite{butz2001hybrid}, automated guided vehicle (AGV) coordination in logistics warehouses \cite{li2022route}, and firefighter tracking in hazardous zones critically depend on accurate position estimation, where traditional methods fail \cite{chen2022tutorial}. In these scenarios, the signal source is often non-cooperative, which lacks structured pilot signals or explicit identifiers, and the dominant propagation is non-line-of-sight (NLoS), due to occlusions by walls, machinery, or debris \cite{dardari2021nlos}. Under such conditions, mainstream localization techniques—including global positioning systems (GPS) and vehicle-to-everything (V2X) protocols—become ineffective, as they assume either the presence of a direct signal path or prior knowledge of the transmitter's behavior. For instance, GPS is unreliable in urban canyons, underground tunnels, and smoke-filled buildings, while cooperative solutions based on roadside units (RSUs) or anchor nodes struggle in dynamic or adversarial environments \cite{sanders2020localizing,huang2024toward,dardari2021nlos}. These limitations expose a significant gap in existing positioning frameworks: the inability to passively localize the unknown sources in geometrically complex and signal-degraded regions. Addressing this challenge is essential not only for safety-critical applications but also for the broader realization of resilient, intelligent, and autonomous communication systems in 6G and beyond.

\begin{figure*}
    \centering
    \includegraphics[width=1\linewidth]{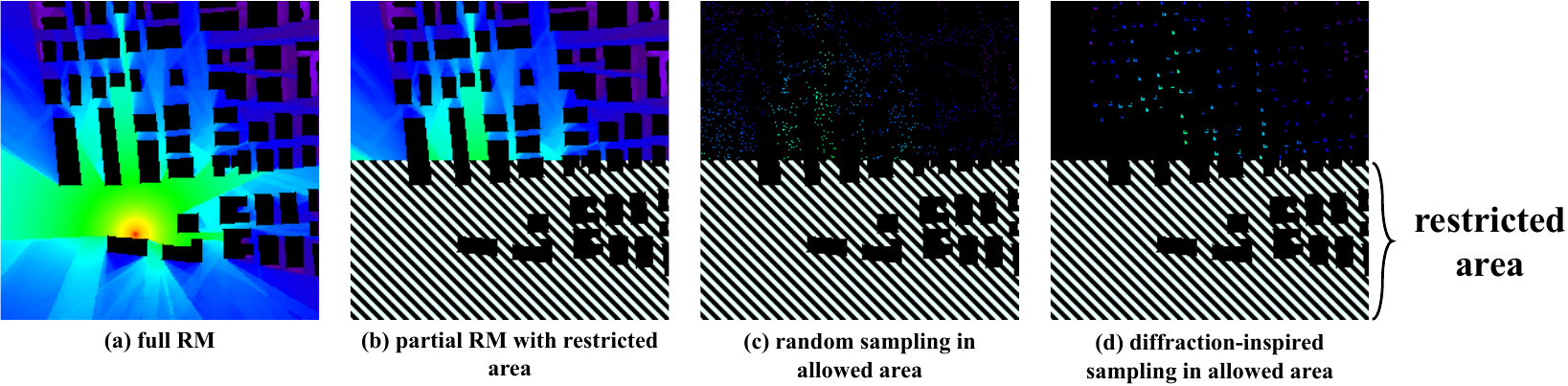}
    \caption{The illustration of the NLoS localization based on RSS information in the allowed area. (a) shows the full RM where the radiation source can be easily obtained from the RM; (b) is the partial RM where the RSS information in the restricted area is unknown; (c) is the random sampling result in the allowed area; and (d) is the diffraction inspired sampling method.}
    \label{fig-sampling-demo}
\end{figure*}

Despite its critical importance, accurate localization of non-cooperative emitters in NLoS environments remains an open and fundamentally ill-posed problem \cite{lin2016enhanced}. The core difficulty lies in two coupled limitations: (i) the absence of tractable priors describing electromagnetic (EM) wave propagation in complex environments \cite{jones2013theory}, and (ii) the inability to perform direct measurements in regions where the target emitter resides \cite{chen2022tutorial}, which are shown in Fig.~\ref{fig-sampling-demo}. First, unlike LoS scenarios where signal behavior can be approximated with simple geometric models, NLoS propagation is governed by complex physical interactions, including diffraction, reflection, and scattering, across heterogeneous surfaces and materials \cite{deschamps1972ray}. These interactions cause nonlinear, spatially irregular signal distributions that defy analytical modeling. As a result, classical Bayesian approaches, such as maximum a posteriori (MAP) estimation, become ill-defined due to the lack of a meaningful prior distribution over the signal field \cite{vapnik1998statistical}. Second, in practical deployments, measurement access is often limited by environmental constraints: the emitter may be located inside a collapsed building, a restricted area, or a dynamically changing zone, rendering direct sampling infeasible \cite{dardari2021nlos}. This is further compounded by the non-cooperative nature of the source, which emits passively and lacks synchronization or structured pilots. In such cases, the only available information is coarse-grained received signal strength (RSS) measurements \cite{liu2015rss}, which are severely distorted by multipath propagation and sensitive to both geometry and material properties. Moreover, without knowledge of the transmitter’s power, even the scale of RSS values is ambiguous, further complicating inference. Finally, the lack of explicit propagation paths means that no principled sampling strategy exists to determine where and how to collect measurements for reliable localization. Existing methods often resort to uniform or heuristic-based sampling, which fails to capture the critical variations induced by environmental geometry, resulting in poor accuracy and high deployment cost. These challenges collectively underscore the need for a fundamentally new paradigm that can handle uncertain physics, sparse observations, and adversarial conditions in a unified and scalable manner.

To overcome the lack of analytical models for signal propagation in NLoS scenarios, the concept of a radio map (RM) has emerged as a powerful alternative for enabling data-driven inference \cite{zeng2024tutorial,wang2025radiodiff,wang2025radiodiff3d,wang2024radiodiff,wang2025radiodiffinv}. An RM captures the spatial distribution of wireless signal characteristics—such as pathloss or received signal strength—across a geographical region, effectively embedding the influence of environmental geometry, material properties, and electromagnetic interactions into a dense signal field \cite{zeng2021toward}. From a probabilistic standpoint, each RM can be interpreted as a sample from an implicit prior distribution that governs how electromagnetic energy propagates in a specific environment \cite{vapnik1998statistical,wang2025radiodiffinv}. This perspective opens the door to recasting the localization problem as a posterior inference task, where sparse RSS measurements serve as observations, and the RM acts as a prior for the otherwise intractable Bayesian prior. If this prior distribution could be explicitly modeled or sampled from, it would enable principled estimation of the emitter’s position, even in the absence of LoS paths. However, RMs are inherently high-dimensional, nonparametric structures, whose internal dependencies are shaped by complex environmental factors \cite{levie2021radiounet}. As such, they lack closed-form parametric representations, making them incompatible with conventional inference techniques that require tractable priors. Furthermore, because each RM is specific to its underlying environment, it is essential to develop methods that can learn and generalize the prior distribution over RMs, conditioned on observable environmental features such as building layouts or base station positions \cite{wang2024radiodiff}. This motivates the use of generative models, which are capable of learning high-dimensional data distributions and capturing the latent structure of signal fields in a flexible, environment-adaptive manner \cite{LDM}. In this way, the RM transitions from a static visualization tool to an active component of the inference pipeline—serving as the learned, physically grounded prior that enables accurate NLoS localization from sparse and noisy measurements.

While learning-based methods can reconstruct the radio map from sparse observations, their performance is fundamentally constrained by the quality and informativeness of the sampled measurements \cite{wang2025radiodiffinv}. In NLoS environments, where signal behavior is highly sensitive to geometry, the RSS samples taken often matter more than how many are taken. To address this, we draw inspiration from knife-edge diffraction theory, which reveals that electromagnetic energy tends to concentrate and vary sharply near geometric discontinuities—particularly building edges and vertices \cite{deschamps1972ray}. These regions exhibit high sensitivity to emitter position and orientation, making them ideal candidates for sampling under information-theoretic criteria such as Fisher information and mutual information with respect to the unknown source. Based on this physical insight, we propose two geometry-aware sampling strategies: a surface-based scheme, which samples along structural boundaries, and a vertex-based scheme, which targets architectural corners where field variation is most pronounced. Unlike heuristic or uniformly random sampling, which can miss critical spatial features, our approach places sensors at locations where RSS values are not only more informative, but also more stable under power normalization, enabling inference without knowledge of absolute transmission power. Remarkably, the number of required samples in our approach scales not with the size of the environment or the number of grid points, but rather with its geometric complexity—e.g., the number of vertices—yielding substantial gains in sampling efficiency. In low-complexity settings, accurate localization and RM reconstruction can be achieved with fewer than 1\% of all spatial positions measured. This physics-informed sampling paradigm transforms sparse measurement collection from an arbitrary design choice into a principled component of the inference framework, enabling practical deployment in real-world NLoS settings with strict sensing constraints.

Motivated by these challenges and opportunities, this paper proposes RadioDiff-Loc, a novel generative localization framework that integrates conditional diffusion models with physics-guided sampling to enable accurate, sampling-efficient localization of non-cooperative NLoS emitters. Unlike conventional approaches that rely on fully sampled observations or handcrafted priors, RadioDiff-Loc learns the statistical structure of radio maps directly from environmental layouts and sparse RSS inputs, enabling end-to-end inference under severe sensing constraints. To the best of our knowledge, this is the first work to realize dense radio map reconstruction and precise emitter localization using only a limited number of uncalibrated RSS measurements from restricted sensing regions, without access to transmitter power or LoS paths. The main contributions of this paper are summarized as folows.
\begin{enumerate}
    \item We propose the RadioDiff-Loc framework to leverage conditional denoising diffusion models for non-cooperative NLoS emitter localization. It reconstructs dense radio maps and estimates emitter positions using only sparsely sampled RSS data from restricted regions—without requiring LoS paths, structured pilot signals, or known transmit power.
    \item To address the challenge of unknown transmission power, we introduce a normalization scheme that enables the diffusion model to generate relative RSS distributions. This allows robust inference and localization from uncalibrated measurements, making the system agnostic to source power.
    \item Inspired by knife-edge diffraction theory, we design two geometry-aware sampling strategies—surface-based and vertex-based—that place sensors at locations with maximal information gain. These strategies reduce the required sampling budget by orders of magnitude and scale with environmental complexity rather than emitter location.
    \item By generating dense, physically plausible radio maps, RadioDiff-Loc enables seamless integration with traditional RSS-based techniques such as fingerprinting and trilateration, bridging generative modeling with legacy localization pipelines for enhanced robustness and accuracy.
\end{enumerate}

\section{Related Works and Preliminaries}
\subsection{Relateds Works}
Recent research in NLoS localization has pursued multiple directions to mitigate the challenges posed by multipath propagation and biased measurements, which remain the predominant sources of error in time-based indoor and urban localization systems \cite{yu2018novel,klus2024robust}. One prominent strand of work reformulates NLoS handling as a pattern recognition problem, leveraging deep learning to classify or regress signal attributes directly from raw observations \cite{bregar2018improving,kim2022uwb,zhao2022nlos}. Early work by Bregar and Mohorčič demonstrated that convolutional neural networks (CNNs) trained on raw ultra-wideband (UWB) channel impulse responses could effectively infer link state and residual range bias \cite{bregar2018improving}, thereby enabling corrected range measurements that, when passed through a classical weighted least-squares solver, achieved over 70\% reduction in median localization error. Building on this foundation, Zhao and Wang employed generative adversarial networks to augment class-imbalanced coal mine datasets \cite{zhao2022nlos}, improving CNN-based link classification accuracy to 91.2\% and realizing measurable localization gains in field deployments. More recent developments have introduced richer input representations—such as channel-state matrices or CSI-derived images—that allow uncertainty estimation and top-K candidate outputs, laying the groundwork for reliable NLoS handling in safety-critical environments.

A second line of research incorporates lightweight learning modules within probabilistic filtering frameworks to improve resilience under mixed LoS/NLoS conditions \cite{cheng2020indoor,cheng2023uwb,kang2025nmap}. Cheng et al. integrated a directional probabilistic data-association gate with an adaptive particle filter to reject spurious ranges and refine tracking performance \cite{cheng2020indoor}; similarly, Kang et al. proposed a novel pipeline, where hypothesis testing segregates LoS and NLoS links, and a shallow neural network adaptively corrects estimates from parallel Kalman and unscented Kalman filters \cite{yong2020robust}. Both approaches underline a key insight: combining explicit link-state discrimination with adaptive filter re-weighting consistently outperforms either strategy alone. In contrast to pre-classification-based methods, optimization-driven techniques have emerged that aim to be intrinsically robust to biased measurements \cite{marano2010nlos,liu2021nlos}. Liu et al. cast the UWB localization problem as a max–min optimization under a correntropy-based loss function, solved using a continuous-time neurodynamic architecture \cite{liu2021nlos}. Notably, this method avoids reliance on link-state labels or statistical bias models and still achieves superior performance in bias-dominated regimes. These developments signal a shift toward robust, bias-invariant estimation objectives as an alternative to pre-screening measurements.

More recently, geometric learning approaches have sought to model multipath explicitly by embedding ray-based physical constraints \cite{wang2020nlos,han2025rayloc}. In RayLoc \cite{han2025rayloc}, Han et al. formulate localization as an inverse ray tracing problem, optimizing both transmitter coordinates and environmental geometry via a fully differentiable simulator. The integration of Gaussian-kernel convolution mitigates gradient sparsity, enabling convergence to centimeter-level accuracy and outperforming conventional CSI-based methods. This hybrid treatment of geometry and learning exemplifies a growing trend toward physically informed neural inference in wireless positioning. Although these lines of research represent significant progress, they also expose critical limitations. Many learning-based models are tightly coupled to specific input modalities or hardware (e.g., UWB) \cite{yu2018novel,schroeder2007nlos}, while optimization frameworks often suffer from high computational cost or require detailed geometric priors. Moreover, both approaches generally assume access to extensive measurement data, which is impractical in scenarios involving non-cooperative sources or inaccessible environments. In contrast, our work takes a fundamentally different approach by recasting the NLoS localization problem as a generative inference task, where the unknown radiation source location and signal distribution are sampled from a learned conditional distribution. By leveraging diffusion-based generative modeling, we learn the spatial prior distribution of electromagnetic signal propagation from environmental geometry and then perform posterior inference based on sparse RSS observations. This framework not only eliminates the need for pilot signals or pre-classified measurements but also offers theoretical guarantees on posterior recoverability under sparse sampling guided by diffraction-informed heuristics. In doing so, our method bridges the gap between data-driven modeling and physics-aware inference, offering a scalable and generalizable solution for high-precision NLoS localization in real-world, measurement-limited scenarios.

\begin{figure*}
    \centering
    \includegraphics[width=1\linewidth]{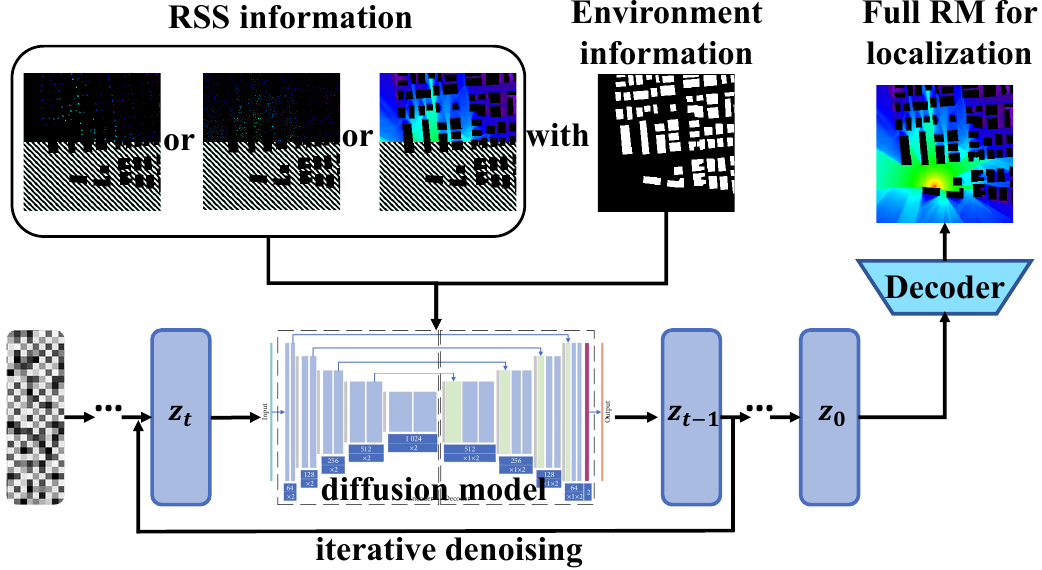}
    \caption{Illustration of the RadioDiff-Loc framework. The RSS information in the allowed areas and the full environment information are input to the DM as the condition to generate the full RM for NLoS localization.}
    \label{fig-system}
\end{figure*}

\subsection{Score-Based Diffusion Model}
DMs have recently gained prominence as powerful generative frameworks capable of modeling complex data distributions and synthesizing high-fidelity samples \cite{ho2020denoising}. Among these, score-based diffusion models define a stochastic generative process through stochastic differential equations (SDEs), where data is gradually perturbed and then iteratively recovered by learning the gradient of the log data density—commonly referred to as the score function \cite{song2020score}. Unlike traditional DMs, which rely on discrete-time Markov chains to simulate the forward and reverse noise processes, score-based models operate in continuous time, rendering them particularly well-suited for inverse problems such as radio map (RM) reconstruction through Bayesian sampling. The forward diffusion process is described by the SDE ad follows.
\begin{align}
d\bm{x} = f(\bm{x}, t) dt + g(t) d\bm{w},
\end{align}
where $f(\bm{x}, t)$ is the drift term, $g(t)$ denotes the time-varying diffusion coefficient, and $d\bm{w}$ is the standard Wiener process. As time progresses, the distribution $p_t(\bm{x})$ converges to an isotropic Gaussian. Recovering the original data entails solving the reverse-time SDE as follows,
\begin{align}
d\bm{x} = \left[ f(\bm{x}, t) - g^2(t) \nabla_{\bm{x}} \log p_t(\bm{x}) \right] dt + g(t) d\bm{\bar{w}},
\end{align}
where the score function $\nabla_{\bm{x}} \log p_t(\bm{x})$ is typically estimated via a neural network $s_\theta(\bm{x}, t)$ trained to approximate:
\begin{align}
s_\theta(\bm{x}, t) \approx \nabla_{\bm{x}} \log p_t(\bm{x}).
\end{align}

For deterministic generation, the equivalent probability flow ordinary differential equation (ODE) eliminates stochasticity as follows.
\begin{align}
d\bm{x} = \left[ f(\bm{x}, t) - \frac{1}{2} g^2(t) \nabla_{\bm{x}} \log p_t(\bm{x}) \right] dt,
\end{align}
providing a continuous analog to the denoising diffusion probabilistic model (DDPM). In DDPM, the forward process is discretized as follows.
\begin{align}
q(\bm{x}_t | \bm{x}_{t-1}) = \mathcal{N}(\bm{x}_t; \alpha_t \bm{x}_{t-1}, \beta_t \bm{I}),
\end{align}
With the score function computed from a learned denoiser $\epsilon_\theta$ as follows.
\begin{align}
s_\theta(\bm{x}, t) = -\frac{\epsilon_\theta(\bm{x}, t)}{\sqrt{1 - \bar{\alpha}_t}}.
\end{align}
This formulation establishes DDPM as a discrete approximation to the continuous score-based paradigm using a variance-preserving mechanism.

In the domain of radio map construction, decoupled diffusion models (DDMs) have been introduced to improve training stability and sample quality, particularly under the RadioDiff framework \cite{huang2024decoupled}. Unlike conventional DMs, which inject noise directly onto the input data, DDMs employ a two-phase process: the signal is first attenuated to a near-zero baseline, followed by the injection of Gaussian noise. The forward process from $\bm{x}_0$ to $\bm{x}_t$ is defined as follows
\begin{align}
q\left(\bm{x}_t \mid \bm{x}_0\right) = \mathcal{N}\left(\gamma_t \bm{x}_0, \delta_t^2 \bm{I}\right),
\end{align}
with time-dependent coefficients $\gamma_t$ and $\delta_t$ controlling the decay of signal content and the variance of injected noise, respectively. The SDE can model this process as follows.
\begin{align}
&d \bm{x}_t = \bm{f}_t \bm{x}_t dt + g_t d\bm{\epsilon}_t,\\
&\bm{f}_t = \frac{d \log \gamma_t}{dt},\\
&\int_{0}^{t}\bm{f}_t dt = -x_0\\
&g_t^2 = \frac{d \delta_t^2}{dt} - 2 \bm{f}_t \delta_t^2,
\end{align}
where $\bm{f}_t$ modulates the deterministic shrinkage and $g_t$ accounts for time-varying stochasticity. To reconstruct the original data, the reverse-time process is solved as follows.
\begin{align}
d \bm{x}_t = \left[\bm{f}_t \bm{x}_t - g_t^2 \nabla_{\bm{x}} \log q\left(\bm{x}_t\right)\right] dt + g_t d\overline{\bm{\epsilon}}_t.
\end{align}
The deterministic transformation to the zero state allows the forward transition to be simplified as follows.
\begin{align}
q(\bm{x}_t|\bm{x}_0) = \mathcal{N}\left(\bm{x}_0 + \int_0^t \bm{f}_\tau d\tau, t\bm{I}\right),
\end{align}
The reverse update step is derived as follows.
\begin{align}
q\left(\bm{z}_{t-\Delta t} \mid \bm{z}_t, \bm{z}_0\right)  &=\mathcal{N}\left(\bm{z}_{t} +\int_t^{t-\Delta t} \bm{f}_t \mathrm{~d} t\right. \notag\\
& \left.\qquad\qquad-\frac{\Delta t}{\sqrt{t}} \bm{\epsilon}, \frac{\Delta t(t-\Delta t)}{t} \bm{I}\right).\label{ddm-reverse}
\end{align}
Through this decoupled perturbation strategy, DDMs enhance generative robustness and efficiency, making them particularly effective for high-resolution RM synthesis in dynamic wireless scenarios.

\section{System Model and Problem Formulation}
We consider a two-dimensional environment denoted as a plane $\mathcal{S} \subset \mathbb{R}^2$, populated with multiple static buildings and a single radiation source. The radiation source, which emits an isotropic spherical wave in steady state, is located at an unknown position $\bm{d} \in \mathcal{S}_r \subset \mathcal{S}$, where $\mathcal{S}_r$ denotes the restricted region of the plane. The remaining area, denoted by $\mathcal{S}_s = \mathcal{S} \setminus \mathcal{S}_r$, constitutes the sensing region, where sensors can be deployed to collect measurements of the RSS. The spatial distribution of buildings is represented by a binary matrix $\bm{H} \in \{0,1\}^{N \times N}$, where each element $H_{i,j} = 1$ indicates the presence of a building at the corresponding grid location $(i,j)$, and $H_{i,j} = 0$ otherwise. All buildings are assumed to be homogeneous in material composition and share the same vertical height $h$, which introduces uniform NLoS effects in the signal propagation model. These buildings act as occluding structures that attenuate and diffract the emitted signal, thereby influencing the spatial distribution of the observed RSS field. To perform localization, a subset of sensor locations $\bm{r} = \{\bm{r}_1, \bm{r}_2, \ldots, \bm{r}_M\} \subset \mathcal{S}_s$ is selected, and the corresponding RSS values $\bm{y} = \{y_1, y_2, \ldots, y_M\}$ are measured. These sampled RSS measurements provide the only observable clues regarding the source location $\bm{d}$, as no direct access to $\mathcal{S}_r$ is permitted. The goal is to accurately estimate the unknown emitter position based on these sparse, indirect measurements. To this end, we propose to train a neural network $\mu_{\theta}$ parameterized by $\theta$, which takes the sampled sensor positions $\bm{r}$ and their corresponding RSS values $\bm{y}$ as input and predicts the location estimate $\hat{\bm{d}}$. The learning process is data-driven and implicitly captures the complex spatial interactions between environmental obstacles, signal attenuation, and multipath propagation. The localization task can be formulated as the following optimization problem as follows
\begin{align}
&\min_{\hat{\bm{d}}} \|\bm{d}-\hat{\bm{d}}\|_{2}+\alpha \|\bm{r}\|_{0},\label{obj}\\
&s.t. \quad \bm{d}\in \mathcal{S}_{r}\tag{\ref{obj}a},\label{c1}\\
&\qquad\;\, \mathcal{S}_{s} = \mathcal{S} / \mathcal{S}_{r}\tag{\ref{obj}b},\label{c2}\\
&\qquad\;\, \bm{r} \in \mathcal{S}_{s}\tag{\ref{obj}c},\label{c3}\\
&\qquad\;\,  \hat{\bm{d}} = \mu_{\theta}(\bm{r}, \bm{y})\tag{\ref{obj}d},\label{c4}
\end{align}
where $\alpha$ is a weighting factor. The objective \eqref{obj} aims to minimize the Euclidean distance between the estimated position $\hat{\bm{d}}$ and the true emitter location $\bm{d}$ while minimizing the number of sampling points. Constraint \eqref{c1} ensures that the emitter lies within the restricted region $\mathcal{S}_r$. Constraint \eqref{c2} defines the sensing region $\mathcal{S}_s$ as the complement of $\mathcal{S}_r$ within the environment $\mathcal{S}$. Constraint \eqref{c3} specifies that all sensor positions must be placed within the accessible sensing region $\mathcal{S}_s$. Finally, constraint \eqref{c4} models the inference process: the neural network $\mu_{\theta}$ maps the spatial distribution of sensor placements and their associated RSS readings to a prediction of the radiation source’s location.

\section{Knife Edge Diffration Inspired Method}
\subsection{Knife Edge Diffration Based Sampling Method}
In the context of NLoS localization, one of the primary challenges lies in accurately estimating the signal field with a limited number of measurements. Due to occlusions and multipath propagation, especially in urban or indoor environments, directly observing the source or relying on line-of-sight propagation is often impossible. A naive uniform sampling strategy in such environments is both inefficient and suboptimal, as most regions contribute marginal information due to diffraction-induced signal decay. To address this, we ground our sparse sampling strategy in physical electromagnetic theory, specifically leveraging the canonical solution of knife-edge diffraction, to identify sampling points that carry maximal information about the underlying field.

Consider a perfectly conducting half-plane occupying $x<0$ with a sharp edge aligned along the $y$-axis. A monochromatic plane wave of wavelength $\lambda$ and incidence angle $\theta_0$ impinges on the edge, inducing a diffracted field that propagates into the shadow region. The incident field is expressed as follows.
\begin{align}
    E^{\text{inc}}(x,z)=E_0\,e^{-jk(x\cos\theta_0 + z\sin\theta_0)}, \quad k = \frac{2\pi}{\lambda},
\end{align}
where the total field satisfies the homogeneous Helmholtz equation with boundary conditions enforced along the conductor, this classical setup, solved rigorously by Sommerfeld, yields a closed-form solution for the field at any shadow-zone point $P(x,z)$ as follows.
\begin{align}
    \frac{E_y(P)}{E_0} = \frac{1 + j}{2}\left[ C(\nu) + j S(\nu) \right], 
\end{align}
where $\nu$ is a dimensionless diffraction parameter determined by the geometry as follows.
\begin{align}
    \nu = \frac{h\sqrt{2(d_1 + d_2)}}{\sqrt{\lambda d_1 d_2}}.
\end{align}
where $h$ is the knife-edge blocking height, $d_1$ and $d_2$ are the distances from the source and receiver to the edge, and $C(\nu)$, $S(\nu)$ are the Fresnel integrals as follows.
\begin{align}
    C(\nu) = \int_0^\nu \cos\left( \frac{\pi}{2} t^2 \right)\, dt, \\ S(\nu) = \int_0^\nu \sin\left( \frac{\pi}{2} t^2 \right)\, dt.
\end{align}
In practical systems, the power loss due to edge diffraction is measured in decibels via the excess-loss factor as follows.
\begin{align}
    &L_d(\text{dB}) = 6.9 + 20\log_{10}\left(\sqrt{(\nu - 0.1)^2 + 1} + \nu - 0.1\right), \\
    &\nu > -0.7,
\end{align}
where a compact and accurate empirical fit is adopted by ITU-R standards for wideband channel modeling. This formulation, which originates from the asymptotic behavior of Fresnel integrals, reveals that most of the diffracted energy is concentrated in a narrow region near the knife edge. To operationalize this observation for measurement design, we consider the Kirchhoff–Helmholtz representation of the total field as an integral over the knife-edge contour as follows.
\begin{align}
    E_y(x,z) = \frac{E_0}{2\pi} \int_{-\infty}^{+\infty} \frac{e^{jkr(s)}}{r(s)} [\cos\theta_i(s) + \cos\theta_d(s)] u(s)\, ds,
\end{align}
where $r(s)$ is the distance from edge segment $s$ to the observation point, and $u(s)$ is the unknown boundary field. Applying stationary phase analysis reveals that the dominant contribution to this integral arises from a small neighborhood around the apex $s=z$, with an effective width as follows.
\begin{align}
    \Delta_F \approx \sqrt{\lambda R},
\end{align}
where $R$ is the effective path length. Outside this region, the contribution to the total field becomes negligible as follows.
\begin{align}
    |E_{\text{tail}}| \le \frac{E_0}{\pi k \Delta},
\end{align}
which indicates that sampling beyond a few Fresnel widths provides minimal incremental information.

To formalize the impact of localized sampling, we discretize the edge into segments and model the received signal $\bm{y} \in \mathbb{C}^{N_p}$ from $N_p$ probe positions as as follows.
\begin{align}
    \bm{y} = \bm{K} \bm{u} + \bm{n}, \quad \bm{n} \sim \mathcal{CN}(0, \sigma^2 \bm{I}),
\end{align}
where $\bm{u} \in \mathbb{C}^{N_b}$ are the unknown complex boundary field values on $N_b$ edge segments, and $\bm{K}$ is a Green's function matrix with entries as follows.
\begin{align}
    K_{ij} = \frac{\Delta s_j}{2\pi} [\cos\theta_i^{(ij)} + \cos\theta_d^{(ij)}] \frac{e^{jkr_{ij}}}{r_{ij}}.
\end{align}
From an information-theoretic standpoint, the informativeness of each sample is quantified by the Fisher information matrix $\bm{J} = \sigma^{-2} \bm{K}^H \bm{K}$. Under far-field conditions, the contribution of each segment decays approximately as $1/s_j^2$, indicating that the apex ($s_j=0$) yields maximal curvature of the likelihood function and thus maximal estimator precision as follows.
\begin{align}
    J_{jj} \propto \frac{1}{s_j^2}.
\end{align}
Alternatively, in a Bayesian setting with a Gaussian prior $\bm{u} \sim \mathcal{CN}(0, \bm{C})$, the mutual information between the boundary field and the noisy observations is given as follows.
\begin{align}
    I(\bm{u}; \bm{y}) = \frac{1}{2} \log \det \left( \bm{I} + \frac{1}{\sigma^2} \bm{K} \bm{C} \bm{K}^H \right).
\end{align}
This measure reaches its steepest growth when the column norm $\|\bm{k}_j\|^2$ is maximized, which occurs near the knife edge due to energy concentration. Moreover, mutual information exhibits submodularity, ensuring that greedy placement strategies that prioritize high-curvature locations, such as corners and rooflines, attain near-optimal sampling efficiency.

These theoretical insights inform our practical method, which is instead of uniformly sampling the environment, we propose a targeted sampling scheme in which sparse RSS measurements are taken near high-diffraction regions, such as building edges and corners. These points serve as information-dense anchors to condition our diffusion-based radio map generator. Because each measurement contributes significant curvature to the posterior distribution, a small number of well-placed probes suffices to constrain the inference of the radiation source location. This not only improves localization accuracy but also drastically reduces measurement overhead, enabling efficient deployment in real-world, resource-constrained scenarios.

\subsection{DM-Based NLoS Localization}
The localization of non-cooperative emitters under NLoS conditions can be rigorously formulated as a Bayesian inference problem. Given sparse RSS measurements $\bm{r} = \{r_1, \dots, r_M\}$ collected at sensor locations $\{\bm{r}_1, \dots, \bm{r}_M\} \subset \mathcal{S}_s$ and a known environmental layout $\bm{H}$, the goal is to estimate the unknown emitter position $\bm{d} \in \mathcal{S}_r$ by maximizing the posterior distribution:
\begin{align}
\bm{d}^{\star} = \arg\max_{\bm{d} \in \mathcal{S}_r} p(\bm{d} \mid \bm{r}, \bm{H}) = \arg\max_{\bm{d}} p(\bm{r} \mid \bm{d}, \bm{H}) p(\bm{d}).
\tag{18}
\end{align}
While the spatial prior $p(\bm{d})$ is often assumed uniform, the likelihood $p(\bm{r} \mid \bm{d}, \bm{H})$ is intractable to model analytically due to diffraction, scattering, and multipath effects in realistic environments.

To circumvent this, we treat RMs as implicit samples from the likelihood distribution. Each RM $\mathcal{R}(\cdot; \bm{d}, \bm{H})$ describes the spatial distribution of signal intensity for a given source and environment, thus serving as a high-dimensional realization of $p(\bm{r} \mid \bm{d}, \bm{H})$. Rather than attempting to model this distribution explicitly, we employ a score-based diffusion model to learn it implicitly. The model is pretrained to approximate the conditional distribution as follows.
\begin{align}
\mathcal{R} \sim p_{\theta}(\mathcal{R} \mid \bm{d}, \bm{H}),
\end{align}
by reversing a forward stochastic differential equation (SDE) that gradually corrupts RMs into Gaussian noise, and learning a neural score estimator $s_\theta(\mathcal{R}, t) \approx \nabla_{\mathcal{R}} \log p_\theta(\mathcal{R} \mid \bm{d}, \bm{H})$. To incorporate sparse measurements during inference, we fine-tune the model into a conditional diffusion model as follows.
\begin{align}
\mathcal{R} \sim p_{\theta^\star}(\mathcal{R} \mid \bm{H}, \bm{r}, \bm{y}),
\tag{19}
\end{align}
where $\bm{y} = \{y_1, \dots, y_M\}$ are normalized RSS values sampled at sensor positions. Crucially, building on diffraction theory, we sample only at building vertices, which are known to carry maximal information due to concentrated edge-diffracted energy. This vertex-aware strategy enables substantial measurement sparsity without compromising fidelity.

As transmission power is unknown, we normalize the RSS inputs to remove power bias:
\begin{align}
\tilde{y}_i = \frac{y_i}{\max_j y_j}, \quad \forall i,
\end{align}
converting the RM into a relative pathloss map. Once the conditional model generates a complete RM $\widehat{\mathcal{R}}(\bm{x})$, localization is achieved by selecting the maximum-intensity point:
\begin{align}
\hat{\bm{d}} = \arg\max_{\bm{x} \in \mathcal{S}_r} \widehat{\mathcal{R}}(\bm{x}),
\tag{21}
\end{align}
which approximates the MAP estimate of the emitter location. Repeated sampling yields an ensemble $\{\hat{\bm{d}}^{(k)}\}$, enabling both robust estimation and uncertainty quantification. This framework transforms the intractable NLoS localization problem into a tractable, data-driven posterior inference task. Unlike prior approaches, as is shown in Fig.~\ref{fig-system}, it (i) models $p(\bm{r} \mid \bm{d}, \bm{H})$ implicitly via learned generative priors; (ii) reduces sampling cost through physically grounded, vertex-targeted RSS probes; and (iii) achieves power-invariant localization via a normalized inference pipeline. The result is a scalable and physically informed method for high-accuracy emitter localization in sparse, multipath-rich environments.

\subsection{Model-Enhanced DM for NLoS Localization}
Beyond direct inference of emitter location, the proposed framework offers a valuable dual-driven fusion capability by enabling both data-driven generation and model-driven refinement within a unified pipeline. Through vertex-aware sparse sampling and conditional diffusion inference, our method reconstructs a high-fidelity RM $\widehat{\mathcal{R}}(\bm{x})$, which approximates the full spatial distribution of RSS across the sensing domain $\mathcal{S}_s$. In effect, this process yields a dense RSS field reconstruction from minimal physical measurements, restoring information typically unavailable in non-cooperative NLoS scenarios. Given that the reconstructed RM encodes the RSS at every point $\bm{x} \in \mathcal{S}_s$, the generative output can be seamlessly interpreted as a surrogate measurement field. This perspective allows classical RSS-based localization methods—such as trilateration, maximum likelihood estimation (MLE), and fingerprint matching—to be applied directly to the generated RM formally, if the reconstructed map satisfies the following equation.
\begin{align}
\widehat{\mathcal{R}}(\bm{x}) \approx \mathbb{E}[r(\bm{x}) \mid \bm{H}, \bm{r}, \bm{y}],
\end{align}
then any algorithm that operates on dense RSS inputs can be re-purposed to post-process $\widehat{\mathcal{R}}(\bm{x})$ and produce refined emitter location estimates.

In this setting, our approach serves as a probabilistic signal field emulator, bridging sparse physical sampling with full-resolution RSS modeling. It thus enables a hybrid inference paradigm: the conditional diffusion model captures global structural priors and nonlinear propagation behavior, while traditional localization techniques contribute geometric interpretability and statistical consistency. By fusing these two perspectives, we achieve greater localization accuracy and robustness than either method alone. Moreover, the dual-driven capability provides flexibility in downstream processing: in safety-critical applications, it allows ensemble voting or cross-validation between generative and model-based estimates; in bandwidth-constrained scenarios, it reduces the need for dense real-time sampling by replacing it with learned field extrapolation. In summary, by reconstructing the full-area RSS distribution from sparse, information-rich measurements, the proposed method enables a powerful integration of deep generative modeling and classical signal-based localization, advancing the frontier of NLoS emitter tracking under extreme sensing constraints.

\section{Experiments Results}
\begin{figure*}[ht]
\centering
\begin{adjustbox}{max width=\textwidth}
\begin{tabular}{cccc}
\includegraphics[width=\linewidth]{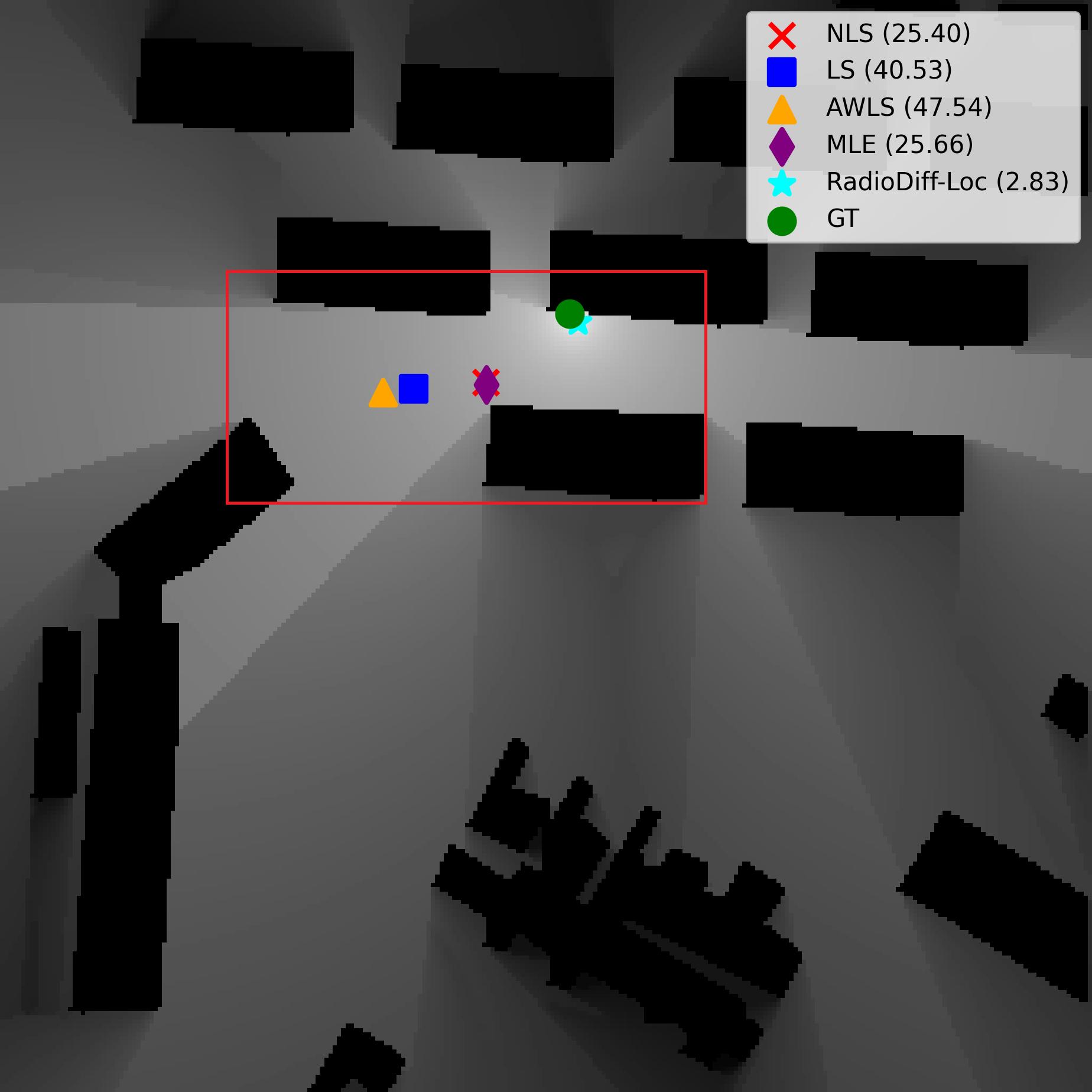} &
\includegraphics[width=\linewidth]{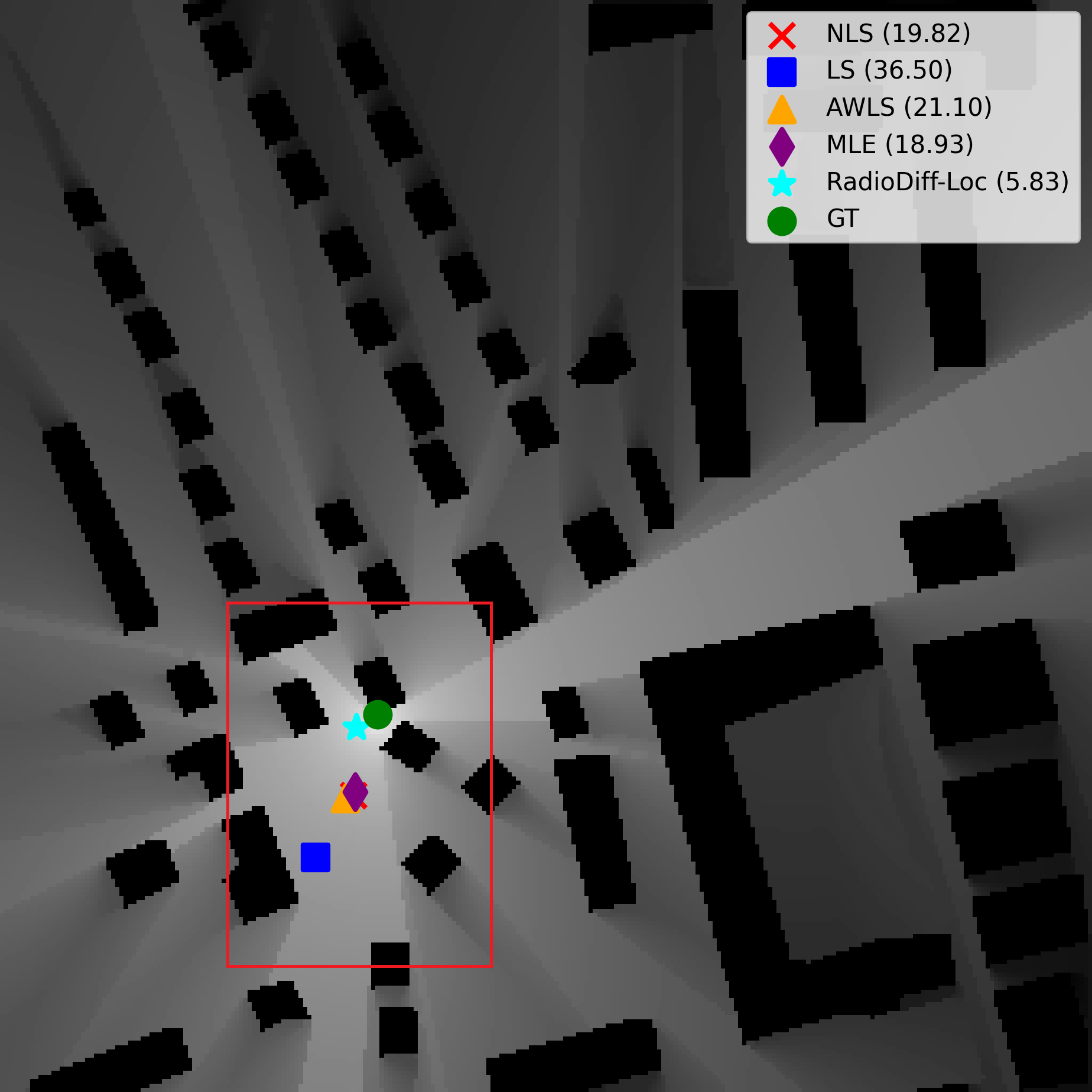} &
\includegraphics[width=\linewidth]{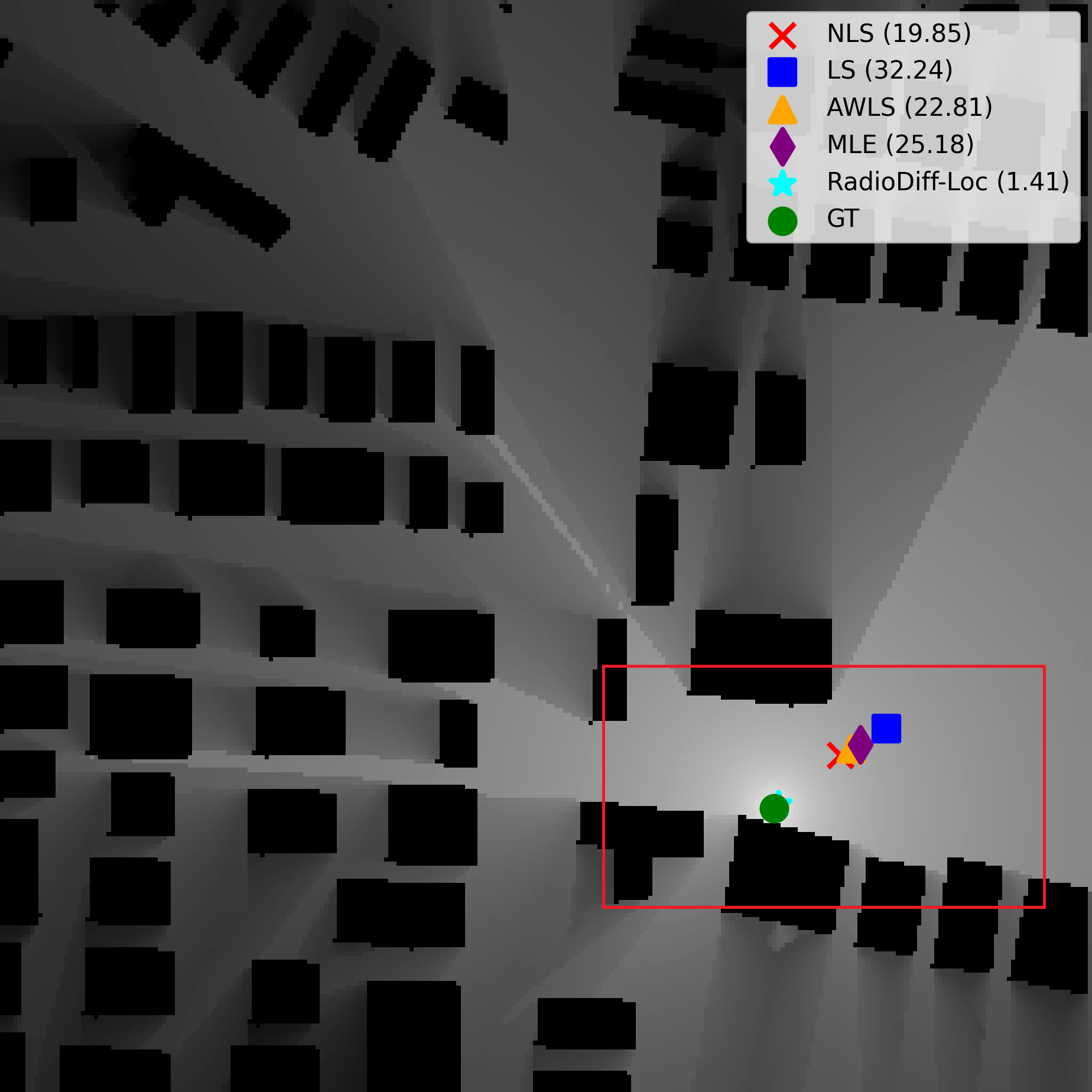} &
\includegraphics[width=\linewidth]{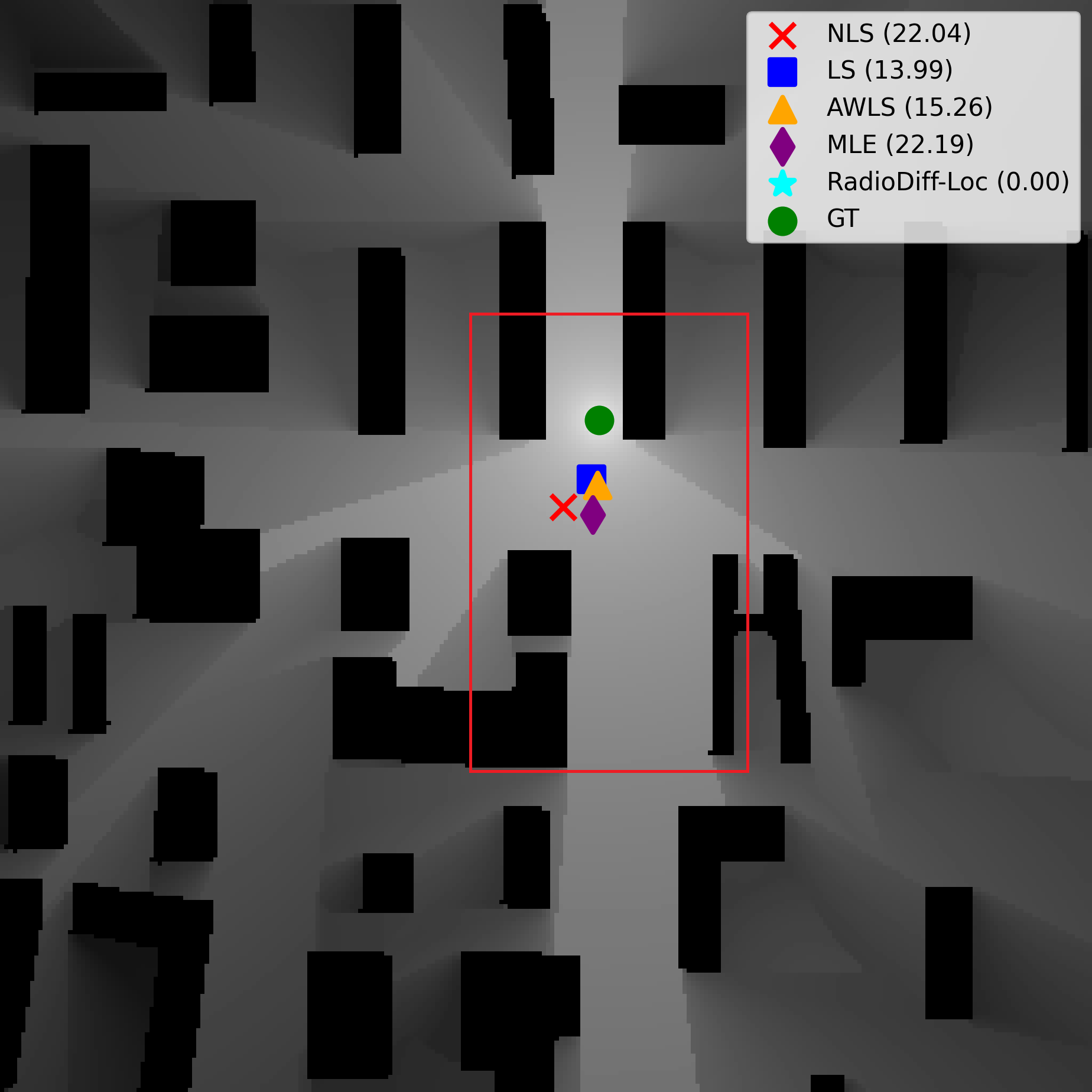} \\
\end{tabular}
\end{adjustbox}

\vspace{1mm}

\begin{adjustbox}{max width=\textwidth}
\begin{tabular}{cccc}
\includegraphics[width=\linewidth]{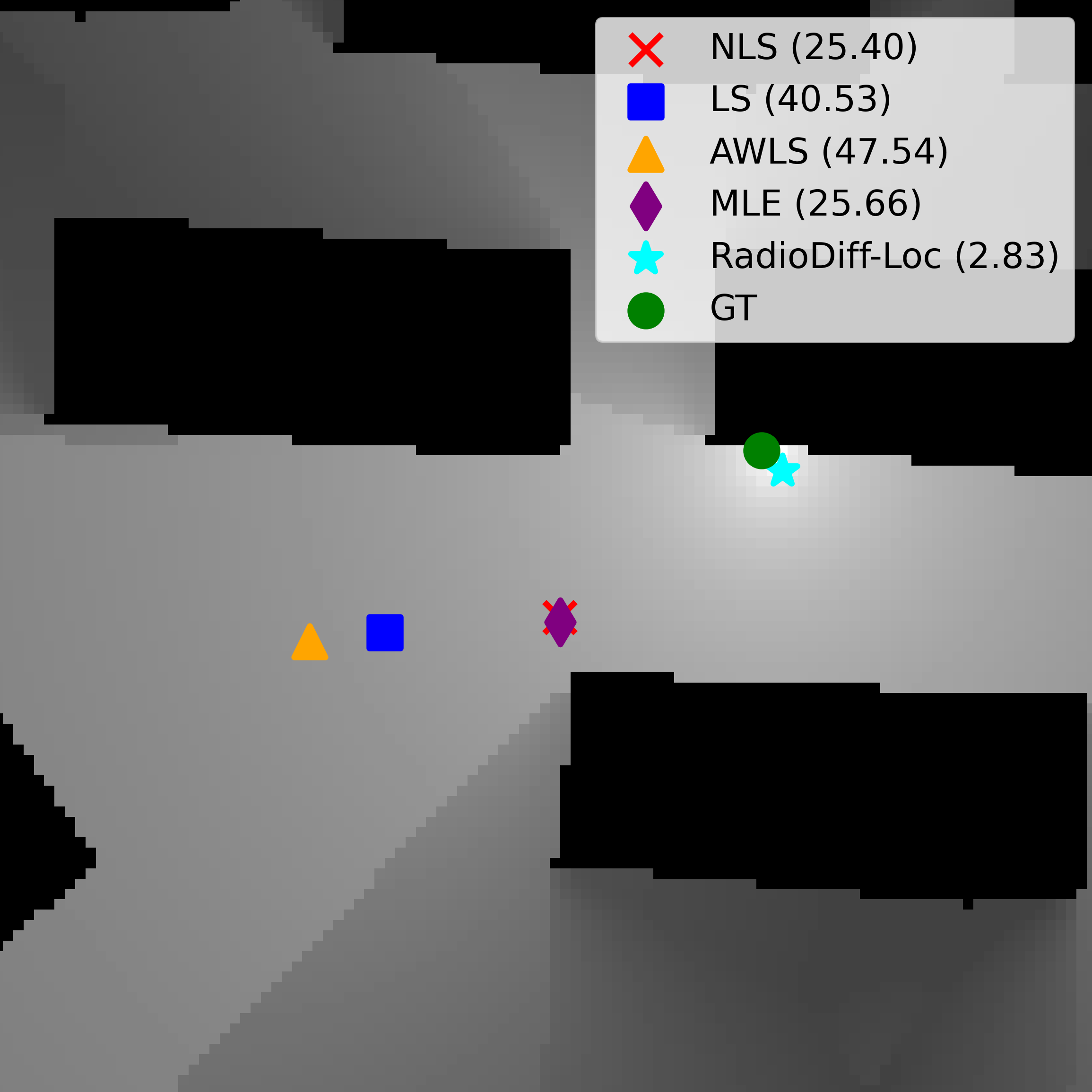} &
\includegraphics[width=\linewidth]{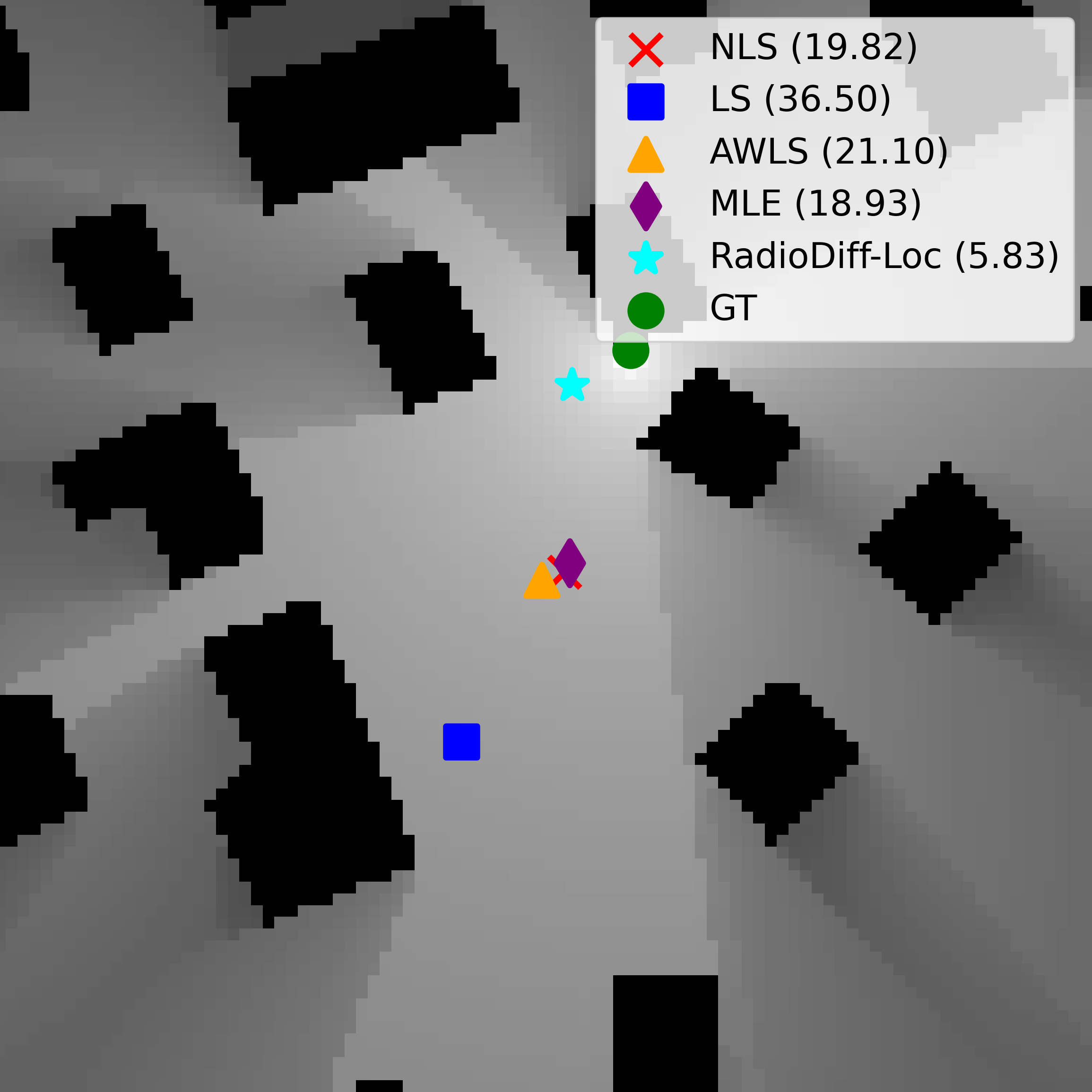} &
\includegraphics[width=\linewidth]{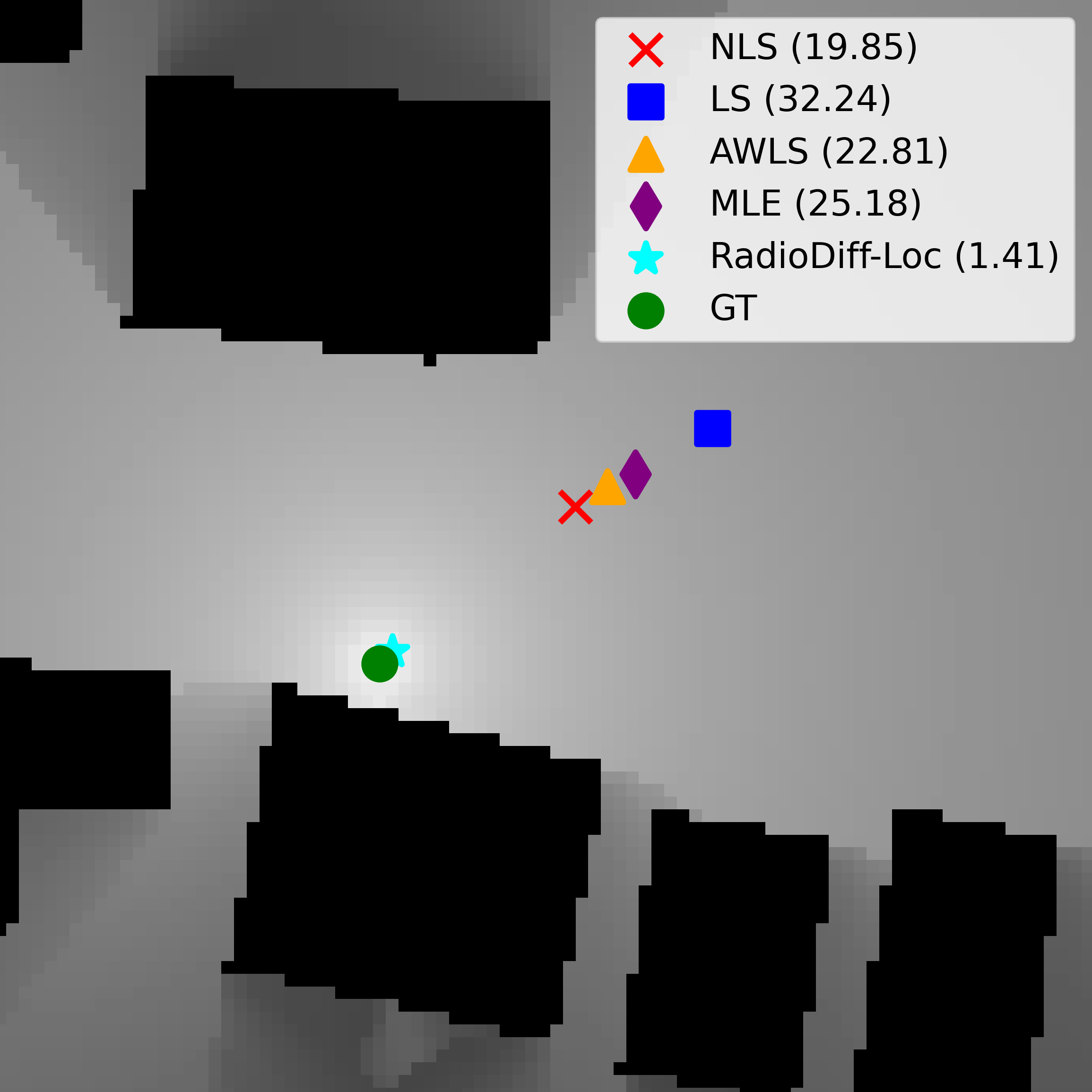} &
\includegraphics[width=\linewidth]{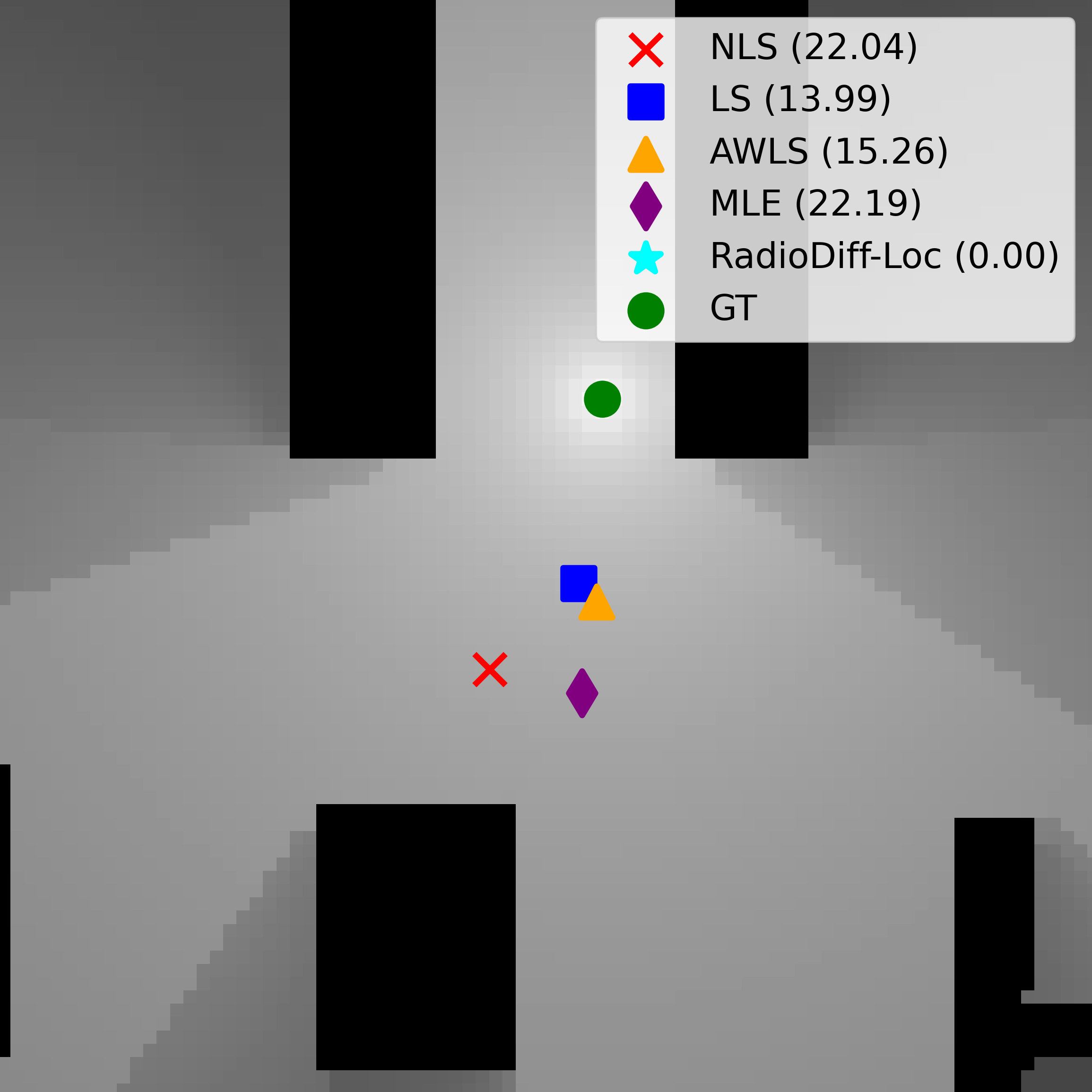} \\
\end{tabular}
\end{adjustbox}

\begin{adjustbox}{max width=\textwidth}
\begin{tabular}{c}
\includegraphics[width=\linewidth]{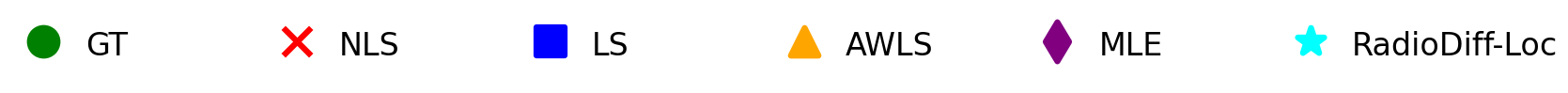} 
\end{tabular}
\end{adjustbox}

\caption{Performance comparison of RSS-based localization methods (NLS, LS, AWLS, MLE) and RadioDiff-Loc across four different environmental scenarios. The top row displays complete localization results with ground truth positions, while the bottom row provides detailed zoomed views of the estimated positions. Each marker represents a different method with corresponding localization errors (in pixels) shown in parentheses.}
\label{fig-srm}
\end{figure*}
\subsection{Dataset}
In this study, we employ the RadioMapSeer dataset \cite{yapar2023first}, to evaluate the performance of the proposed NLoS localization and radio map reconstruction framework. The dataset contains 700 urban-scale radio maps, each associated with 80 distinct transmitter locations and their corresponding ground-truth pathloss measurements. The building layouts are sourced from OpenStreetMap and span six representative metropolitan areas: Ankara, Berlin, Glasgow, Ljubljana, London, and Tel Aviv. Each map includes between 50 and 150 buildings, offering a diverse range of urban morphologies. All radio maps are formatted as 256×256 binary morphological images, where each pixel represents 1 square meter of physical space, with values of 1 indicating building presence and 0 representing open areas. The transmitter and receiver heights are uniformly fixed at 1.5 meters, and building heights are set at 25 meters. Signal transmission is standardized across the dataset, with each transmitter operating at 23 dBm power and a 5.9 GHz carrier frequency. The ground-truth pathloss values are computed by solving Maxwell’s equations, accounting for both reflection and diffraction effects caused by the environment. To ensure fair evaluation and robust generalization, we partition the dataset into 600 training and 100 testing maps, ensuring no geographic overlap between the splits. This comprehensive and high-fidelity dataset enables a rigorous assessment of the proposed method under diverse and realistic propagation scenarios, supporting evaluation across varying levels of environmental complexity and NLoS conditions.

\subsection{Implementation Details}
In this study, the neural network input is formulated as a three-channel tensor \( \mathbf{X} \in \mathbb{R}^{256 \times 256 \times 3} \). The input channels consist of a binary building layout map, which explicitly encodes the spatial distribution of buildings, and two identical sparse-sampled signal strength maps. The sparse signal maps are obtained by applying binary masks that retain measurements exclusively at predefined sampling points, while unsampled locations are set to zero. The repetition of the sparse signal maps across two channels is intended to enhance the feature representation capacity of the network. The network output corresponds to a full-resolution signal strength map \( \mathbf{Y} \in \mathbb{R}^{256 \times 256 \times 1} \), representing the spatial distribution of radio wave intensities throughout the environment. To determine sampling locations, five distinct mask-based strategies are employed, incorporating both geometric priors and randomized sampling to explore their respective impacts on model performance as follows.
\begin{itemize}
    \item \textbf{Random Sampling}: A baseline configuration in which sampling points are uniformly distributed across the environment without regard to geometric features. 
    \item \textbf{Edge Sampling Mask}: According to electromagnetic ray tracing principles, changes in the propagation direction of electromagnetic waves occur primarily at the edges of environmental structures, such as the boundaries of buildings. These edge regions serve as dominant contributors to diffraction and reflection phenomena, and thus carry critical information about the underlying wave interactions. In particular, knife-edge diffraction theory confirms that significant energy concentration and propagation behavior are governed by these structural discontinuities. Motivated by this insight, we adopt an edge sampling strategy, wherein RSS measurements are collected exclusively from points located near building edges. This targeted sampling approach maximizes informational efficiency while minimizing measurement redundancy in low-contribution regions.
    \item \textbf{Vertex Sampling}: Based on knife-edge diffraction theory, electromagnetic waves experience significant diffraction at structural discontinuities, particularly at the vertices of buildings where multiple edges intersect. These vertex regions serve as critical points that strongly influence wave propagation, especially under NLoS conditions. As such, the RSS information captured at building vertices provides high informational value for reconstructing the spatial signal field. Motivated by this, we adopt a vertex sampling strategy, wherein measurements are collected exclusively at or near building corners. This approach concentrates sensing resources at geometrically and physically meaningful locations, enabling accurate radio map reconstruction and emitter localization with minimal sampling overhead.
\end{itemize}
To further validate the effectiveness of our proposed method, we conduct comparative experiments with several classical RSS-based localization algorithms, including least squares (LS) \cite{wang2015linear}, adaptive weighted least squares (AWLS) \cite{zhang2015robust}, maximum Bayesian estimation (MBE) \cite{huan2022indoor}, and nonlinear least squares (NLS) \cite{ng2022kernel}. By guiding the sparse sampling process with explicit structural cues, these mask-based input strategies enable the neural network to prioritize informative spatial regions. This, in turn, enhances the model’s capability in reconstructing non-line-of-sight signal distributions and accurately localizing signal sources. All experiments are conducted on a single NVIDIA GeForce RTX 4090 GPU (49~GB) using CUDA 12.8. The model is trained from scratch using the Adam optimizer with an initial learning rate of \(5 \times 10^{-5}\), which linearly decays to a minimum of \(5 \times 10^{-6}\). A batch size of 48 is used without gradient accumulation. Mixed-precision training is disabled to ensure numerical stability. An exponential moving average (EMA) of model parameters is updated every 10 steps after the first 10000 iterations to improve training stability. All implementations are developed using PyTorch.
\begin{table*}[ht]
\centering
  \caption{Quantitative Comparison of Different Sampling Methods for DM}
  \resizebox{\linewidth}{!}{
    \begin{tabular}{ccccccc}
    \toprule
    \multicolumn{1}{c|}{Method}          & \multicolumn{1}{c|}{NMSE}   & \multicolumn{1}{c|}{RMSE}   & \multicolumn{1}{c|}{SSIM}   & \multicolumn{1}{c|}{PNSR}  & \multicolumn{1}{c|}{LE}    & Sampling Ratio \\ \hline
    \multicolumn{7}{c}{\textit{Edge-Based Sampling}}\\ \hline
    \multicolumn{1}{c|}{Random}          & \multicolumn{1}{c|}{0.0056} & \multicolumn{1}{c|}{0.0219} & \multicolumn{1}{c|}{0.9636} & \multicolumn{1}{c|}{33.50}  & \multicolumn{1}{c|}{2.017} & 2.61\%         \\ \hline
    \multicolumn{1}{c|}{Edge Sampling}   & \multicolumn{1}{c|}{0.0055} & \multicolumn{1}{c|}{0.0212} & \multicolumn{1}{c|}{0.9664} & \multicolumn{1}{c|}{33.80}  & \multicolumn{1}{c|}{1.945} & 2.61\%         \\ \hline
    \multicolumn{7}{c}{\textit{Vertex-Based Sampling}}\\ \hline
    \multicolumn{1}{c|}{Random}          & \multicolumn{1}{c|}{0.0072} & \multicolumn{1}{c|}{0.0232} & \multicolumn{1}{c|}{0.961}  & \multicolumn{1}{c|}{32.96} & \multicolumn{1}{c|}{3.035} & 0.96\%         \\ \hline
    \multicolumn{1}{c|}{Vertex Sampling} & \multicolumn{1}{c|}{0.0069} & \multicolumn{1}{c|}{0.0232} & \multicolumn{1}{c|}{0.9601} & \multicolumn{1}{c|}{32.96} & \multicolumn{1}{c|}{2.770}  & 0.96\%         \\ \bottomrule
    \end{tabular}
}\label{tab-2}
\end{table*}

\subsection{Performance Metrics}
To comprehensively evaluate the quality of radio map reconstruction and the accuracy of emitter localization, we employ the following quantitative performance metrics:
\begin{itemize}
    \item \textbf{Mean Squared Error (MSE)} and \textbf{Root Mean Squared Error (RMSE)}: These metrics quantify the average squared deviation between the predicted radio map \( \hat{I}(m, n) \) and the ground truth map \( I(m, n) \) over all pixels \( (m, n) \). They are defined as:
    \begin{equation}
        \mathrm{MSE} = \frac{1}{NM} \sum_{m=1}^{M} \sum_{n=1}^{N} \left( \hat{I}(m,n) - I(m,n) \right)^2,
    \end{equation}
    \begin{equation}
        \mathrm{RMSE} = \sqrt{\mathrm{MSE}}.
    \end{equation}
    
    \item \textbf{Normalized Mean Squared Error (NMSE)}: This metric measures the relative reconstruction error normalized by the total energy of the ground truth signal:
    \begin{equation}
        \mathrm{NMSE} = \frac{\sum_{m=1}^{M} \sum_{n=1}^{N} \left( \hat{I}(m,n) - I(m,n) \right)^2}{\sum_{m=1}^{M} \sum_{n=1}^{N} I(m,n)^2}.
    \end{equation}
    
    \item \textbf{Structural Similarity Index Measure (SSIM)}: SSIM evaluates perceptual similarity between the predicted and ground truth maps by considering luminance, contrast, and structural information. Given two image patches \( x \) and \( y \), SSIM is defined as:
    \begin{equation}
        \mathrm{SSIM}(x, y) = \frac{(2\mu_x \mu_y + C_1)(2\sigma_{xy} + C_2)}{(\mu_x^2 + \mu_y^2 + C_1)(\sigma_x^2 + \sigma_y^2 + C_2)},
    \end{equation}
    where \( \mu_x, \mu_y \) are local means, \( \sigma_x^2, \sigma_y^2 \) are variances, and \( \sigma_{xy} \) is the covariance between \( x \) and \( y \). Constants \( C_1 = (K_1 L)^2 \), \( C_2 = (K_2 L)^2 \), and \( C_3 = C_2 / 2 \) are included to stabilize the computation.

    \item \textbf{Peak Signal-to-Noise Ratio (PSNR)}: PSNR assesses the ratio between the peak signal power and the power of the reconstruction error. It is calculated as:
    \begin{equation}
        \mathrm{PSNR} = 10 \log_{10} \left( \frac{r^2}{\mathrm{MSE}} \right),
    \end{equation}
    where \( r \) denotes the dynamic range of the signal. PSNR provides a logarithmic scale measure of overall reconstruction quality, particularly sensitive to edge preservation.

    \item \textbf{Localization Error (LE)}: To evaluate emitter localization performance, we compute the Euclidean distance between the predicted emitter position \( \hat{d} \) and the ground truth position \( d \). Averaged over multiple samples, this metric quantifies the model’s localization accuracy:
    \begin{equation}
        \mathrm{LE} = \frac{1}{K} \sum_{k=1}^{K} \| \hat{d}^{(k)} - d^{(k)} \|_2,
    \end{equation}
    where \( K \) is the number of evaluated test instances.
\end{itemize}
Together, these metrics offer a comprehensive evaluation framework, jointly reflecting pixel-wise fidelity, structural quality, perceptual accuracy, and localization precision in challenging NLoS environments.

\subsection{Comparison with Traditional Localization Methods}
\begin{figure}
    \centering
    \includegraphics[width=1\linewidth]{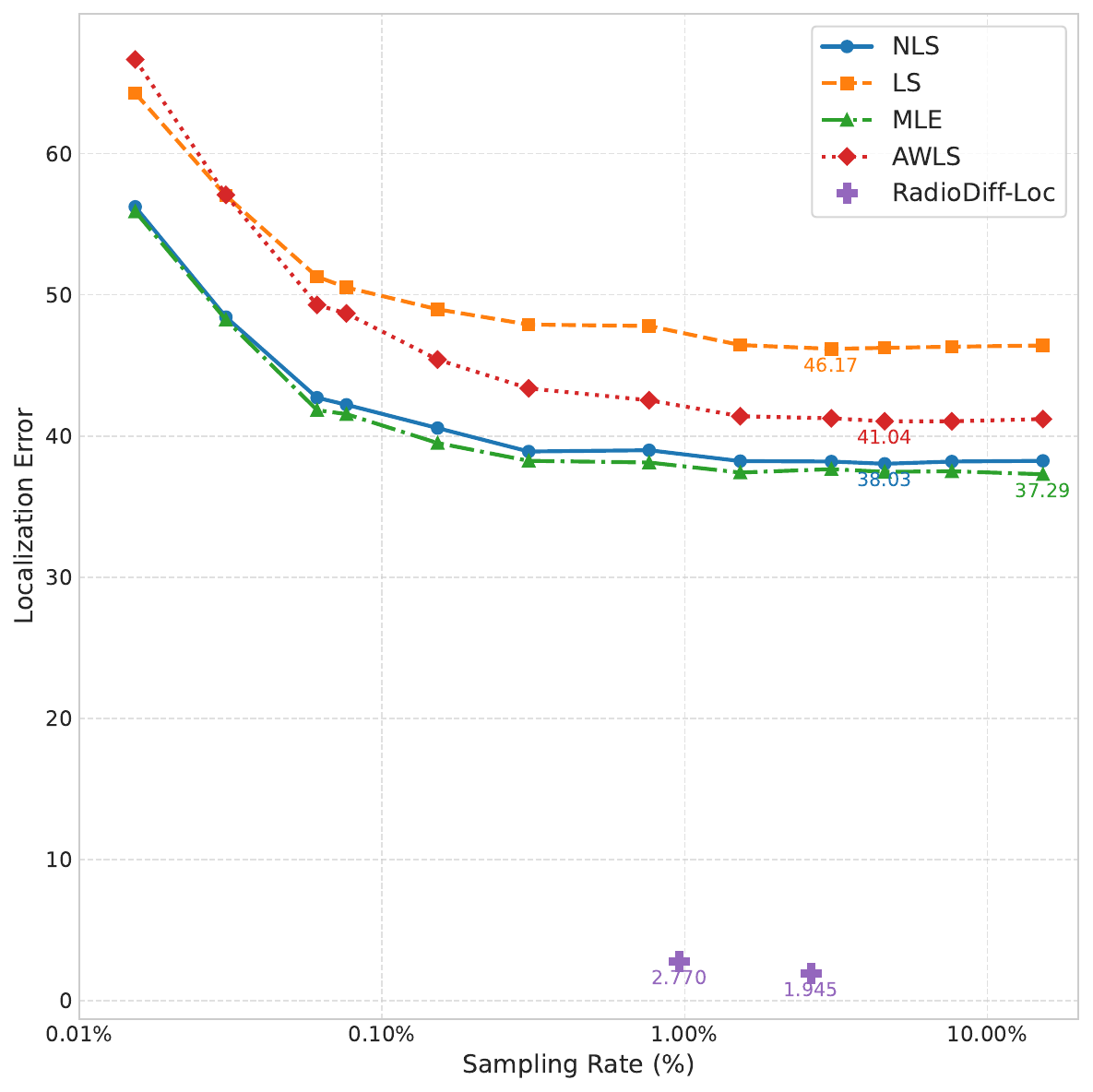}
    \caption{Localization performance of different methods on sampling rates.}
    \label{fig-tradition}
\end{figure}
In the experiments, we evaluate the performance of these methods under a \textbf{global random sampling strategy} with a sampling rate of \textbf{3.05\%}, where observation points are randomly selected across the entire scene. In contrast, RadioDiff-Loc is evaluated under a more challenging setting, where observations are restricted to the \textbf{upper-half region} of the scene with a significantly lower sampling rate. The average localization errors (in meters) of traditional methods and RadioDiff-Loc are summarized in Table~\ref{tab:traditional_methods} and Fig.~\ref{fig-tradition}. Notably, the RadioDiff-Loc result (3.035 m) corresponds to the \textit{random sampling} configuration from the \textbf{Vertex + Random Sampling vs. Random Sampling} experiment. In that setup, the sampling rate is approximately 0.96\%, and random samples constitute about 28.95\% of the total sampling points.

\begin{table}[H]
\centering
\caption{Localization Errors (in meters) of Traditional RSS-Based Methods and RadioDiff-Loc under Different Sampling Strategies}
\label{tab:traditional_methods}
\begin{tabular}{llc}
\toprule
\textbf{Sampling Region} & \textbf{Method} & \textbf{Localization Error (m)} \\
\midrule
\multirow{5}{*}{Global Random} 
    & LS            & 28.43 \\
    & AWLS          & 24.86 \\
    & MBE          & 21.47 \\
    & NLS           & 18.08 \\
\midrule
\multirow{4}{*}{Upper Half Only} 
    & LS   & 38.56 \\
    & AWLS & 41.25 \\
    & MBE & 38.83 \\
    & NLS  & 38.19 \\
\midrule
\multicolumn{2}{l}{RadioDiff-Loc (Upper Half)} & 3.035 \\
\bottomrule
\end{tabular}
\end{table}

As shown in Table~\ref{tab:traditional_methods}, traditional methods suffer from substantial performance degradation, especially when the observation region is limited to the upper half of the scene. Even with a higher sampling rate and access to the entire environment, their localization errors remain significantly larger than those of our method.

In contrast, \textbf{RadioDiff-Loc} consistently achieves \textbf{superior accuracy} despite the lower sampling rate and reduced observation area. Whether employing random, edge-based, or hybrid sampling strategies, RadioDiff-Loc exhibits remarkable \textbf{efficiency}, \textbf{robustness}, and strong generalization capabilities. These results highlight its advantage in capturing spatial signal structures and leveraging incomplete observations far more effectively than traditional approaches.

To complement these quantitative results, Figure~\ref{fig-srm} provides a visual comparison of localization performance across different environments. The visualizations further highlight the substantial gap in accuracy between traditional methods and RadioDiff-Loc, especially under sparse and constrained sampling conditions.

\subsection{Quantitative Analysis of Knife-Edge Diffraction Inspired Sampling}
To evaluate the effectiveness of our proposed sampling strategies, Table~\ref{tab-2} presents the performance of the conditional diffusion model in RM reconstruction and NLoS source localization under three different sampling methods: random sampling, edge-based sampling, and vertex-based sampling. For fair comparison, the number of sampling points in the random strategy is matched to the number of edges or vertices used in the other two methods, ensuring that all approaches operate under identical sampling budgets. Across all test environments, the average sampling rate for edge-based sampling is 2.61\%, while that of vertex-based sampling is significantly lower at only 0.96\%. The results clearly demonstrate that both edge-based and vertex-based strategies outperform random sampling in terms of RM reconstruction quality and localization accuracy. This confirms that sampling points concentrated near building structures, where diffraction and signal variation are most pronounced, provide more informative measurements than those selected randomly across the domain. Notably, the vertex-based strategy achieves a localization error below 3 meters while utilizing less than 1\% of the total grid points, highlighting its exceptional efficiency. In contrast, random sampling not only requires more points to achieve comparable accuracy but also exhibits a steeper degradation in performance under reduced sampling rates. Specifically, when the sampling budget is constrained, random sampling experiences an increase in localization error exceeding 50\%, whereas vertex-based sampling shows a more graceful degradation with only a 40\% increase in error. This result further validates that building vertices serve as high-information regions, supporting our theoretical justification rooted in knife-edge diffraction analysis. Overall, these experiments underscore the practical value of geometry-aware sparse sampling, and confirm that the proposed vertex-based strategy offers an optimal trade-off between measurement cost and localization precision in NLoS environments.

\subsection{Model Enhanced Performance Analysis}
To further exploit the reconstructed RM, we integrate classical RSS-based localization algorithms with our diffusion-generated RMs, forming a dual-driven framework that combines learned priors with model-based refinement. We evaluate three representative methods as follows.
\textbf{Top-$k$ Weighted Centroid}: This method computes the weighted centroid over the top-$k$ highest-intensity pixels in the RM. The estimated position is $(\bar{y}, \bar{x}) = \left( \frac{\sum i \cdot I(i,j)}{\sum I(i,j)}, \frac{\sum j \cdot I(i,j)}{\sum I(i,j)} \right)$, where the sums are taken over the top-$k$ set.
\textbf{Threshold Region Center (Thr RC)}: This method selects all pixels with intensity above a given percentile threshold $T_p$, then computes the geometric center: $(\bar{y}, \bar{x}) = \left( \frac{1}{|R|} \sum i, \frac{1}{|R|} \sum j \right)$, where $(i,j) \in R = \{(i,j) \mid I(i,j) \geq T_p\}$.
\textbf{Largest Blob Centroid (LBC)}: The RM is binarized using a relative threshold $T_r = \alpha \cdot \max I(i,j)$, and the centroid is computed over the largest 8-connected component: $(\bar{y}, \bar{x}) = \left( \frac{1}{|C|} \sum i, \frac{1}{|C|} \sum j \right)$, where $(i,j) \in C$. These post-processing strategies demonstrate that our method not only produces high-quality RMs but also enables effective fusion with traditional RSS methods, realizing a robust dual-driven localization framework.

\begin{table}[h]
\centering
\caption{Performance Comparison of Edge Sampling and Random Sampling}
\label{tab:mask_vs_maskr}
\begin{tabular}{lccc}
\toprule
\textbf{Metric} & \textbf{Edge} & \textbf{Random} & \textbf{Improvement (\%)} \\
\midrule
Top-5 WC       & \textbf{$2.10 \pm 3.50$}  & $2.31 \pm 3.57$  & \textcolor{red}{\textbf{+9.3\%}} \\
Top-10 WC      & \textbf{$2.09 \pm 3.51$}  & $2.30 \pm 3.56$  & \textcolor{red}{\textbf{+9.3\%}} \\
Top-20 WC      & \textbf{$2.09 \pm 3.52$}  & $2.30 \pm 3.55$  & \textcolor{red}{\textbf{+9.3\%}} \\
Top-50 WC      & \textbf{$2.10 \pm 3.50$}  & $2.30 \pm 3.52$  & \textcolor{red}{\textbf{+8.8\%}} \\
Thr 95\% RC    & \textbf{$8.21 \pm 7.38$}  & $8.22 \pm 7.39$  & \textcolor{red}{\textbf{+0.1\%}} \\
Thr 97\% RC    & \textbf{$5.89 \pm 5.59$}  & $5.95 \pm 5.57$  & \textcolor{red}{\textbf{+1.0\%}} \\
Thr 99\% RC    & \textbf{$3.42 \pm 3.89$}  & $3.51 \pm 3.87$  & \textcolor{red}{\textbf{+2.7\%}} \\
Thr 99.5\% RC  & \textbf{$2.70 \pm 3.57$}  & $2.84 \pm 3.55$  & \textcolor{red}{\textbf{+4.8\%}} \\
Thr 99.9\% RC  & \textbf{$2.14 \pm 3.48$}  & $2.33 \pm 3.51$  & \textcolor{red}{\textbf{+8.3\%}} \\
LBC (0.85)     & \textbf{$2.14 \pm 3.48$}  & $2.35 \pm 3.48$  & \textcolor{red}{\textbf{+8.7\%}} \\
LBC (0.9)      & \textbf{$2.04 \pm 3.51$}  & $2.25 \pm 3.54$  & \textcolor{red}{\textbf{+9.0\%}} \\
LBC (0.95)     & \textbf{$2.01 \pm 3.52$}  & $2.22 \pm 3.57$  & \textcolor{red}{\textbf{+9.3\%}} \\
LBC (0.99)     & \textbf{$1.94 \pm 3.58$}  & $2.11 \pm 3.66$  & \textcolor{red}{\textbf{+7.9\%}} \\
\bottomrule
\end{tabular}
\end{table}

\begin{table}[h]
\centering
\caption{Performance Comparison of Edge Sampling and Random Sampling}
\label{tab:vertexr_vs_maskrr}
\begin{tabular}{lccc}
\toprule
\textbf{Metric} & \textbf{Hybrid} & \textbf{Random} & \textbf{Improvement (\%)} \\
\midrule
Top-5 WC       & \textbf{$2.79 \pm 4.20$}  & $2.93 \pm 4.83$  & \textcolor{red}{\textbf{+5.0\%}} \\
Top-10 WC      & \textbf{$2.77 \pm 4.19$}  & $2.91 \pm 4.83$  & \textcolor{red}{\textbf{+4.9\%}} \\
Top-20 WC      & \textbf{$2.77 \pm 4.19$}  & $2.91 \pm 4.82$  & \textcolor{red}{\textbf{+4.8\%}} \\
Top-50 WC      & \textbf{$2.77 \pm 4.18$}  & $2.90 \pm 4.80$  & \textcolor{red}{\textbf{+4.5\%}} \\
Thr 95\% RC    & \textbf{$9.04 \pm 7.74$}  & $8.95 \pm 7.95$  & \textcolor{blue}{\textbf{-1.0\%}} \\
Thr 97\% RC    & \textbf{$6.69 \pm 6.13$}  & $6.62 \pm 6.42$  & \textcolor{blue}{\textbf{-1.2\%}} \\
Thr 99\% RC    & \textbf{$4.06 \pm 4.46$}  & $4.06 \pm 5.00$  & \textbf{0.0\%} \\
Thr 99.5\% RC  & \textbf{$3.34 \pm 4.18$}  & $3.35 \pm 4.77$  & \textcolor{red}{\textbf{+0.4\%}} \\
Thr 99.9\% RC  & \textbf{$2.80 \pm 4.16$}  & $2.91 \pm 4.79$  & \textcolor{red}{\textbf{+3.9\%}} \\
LBC (0.85)     & \textbf{$2.82 \pm 4.13$}  & $2.91 \pm 4.75$  & \textcolor{red}{\textbf{+3.2\%}} \\
LBC (0.9)      & \textbf{$2.75 \pm 4.16$}  & $2.89 \pm 4.80$  & \textcolor{red}{\textbf{+4.9\%}} \\
LBC (0.95)     & \textbf{$2.73 \pm 4.20$}  & $2.90 \pm 4.84$  & \textcolor{red}{\textbf{+5.7\%}} \\
LBC (0.99)     & \textbf{$2.76 \pm 4.24$}  & $2.97 \pm 4.85$  & \textcolor{red}{\textbf{+7.0\%}} \\
\bottomrule
\end{tabular}
\end{table}

To rigorously assess the effectiveness of the proposed geometry-aware sampling strategies, we conducted a series of comparative experiments under fixed sampling budgets. The goal is to evaluate how incorporating geometric priors—specifically, building edges and vertices—impacts localization accuracy, while maintaining comparable performance in RM reconstruction. Table~\ref{tab:mask_vs_maskr} presents the results of Edge Sampling versus baseline Random Sampling. Notably, Top-$k$ Weighted Centroid errors are reduced by approximately 8.8\% to 9.3\% across various $k$, highlighting the advantage of sampling near high-diffraction regions along building boundaries. Similar gains are observed in the Thr RC and LBC metrics, confirming that edge-proximal samples capture critical diffraction and shadowing effects that are informative for emitter localization. Table~\ref{tab:vertexr_vs_maskrr} compares the hybrid Vertex + Random Sampling strategy with pure random sampling. By including strategically placed vertex samples—particularly at building corners and roof edges—the method achieves consistent improvements in localization performance. Specifically, WC errors decrease by 4.5\% to 5.0\%, while LBC accuracy improves by 3.2\% to 7.0\%, with the most significant gain occurring at the 0.99 threshold level. Although most RC metrics remain stable, slight degradations (1.0\%–1.2\%) are observed at the 95\% and 97\% thresholds. Overall, these results demonstrate that geometry-guided sampling, particularly vertex-informed strategies, significantly enhances localization precision without degrading RM reconstruction quality.

\section{Conclusion}

In this work, we proposed RadioDiff-Loc, a novel diffusion-based localization framework designed for non-cooperative signal sources in NLoS environments. By incorporating physical insights from knife-edge diffraction theory, our method introduces geometry-aware sampling strategies that prioritize measurements near building edges and vertices, significantly enhancing localization accuracy with minimal sampling cost. Extensive experiments conducted on realistic urban-scale datasets demonstrate that combining structural priors with conditional generative modeling yields a scalable and robust solution for sparse radio map reconstruction and accurate emitter localization. This framework offers a promising direction for practical deployment in complex, measurement-limited wireless environments.

\bibliography{ref}
\bibliographystyle{IEEEtran}

\ifCLASSOPTIONcaptionsoff
  \newpage
\fi

\end{document}